# Ultralow lattice thermal conductivity and high thermoelectric performance near room temperature of Janus monolayer HfSSe

Jayanta Bera, Atanu Betal and Satyajit Sahu*

Department of Physics, Indian Institute of Technology Jodhpur, Jodhpur 342037, India

**Abstract:**

Two-dimensional transition metal di-chalcogenides (TMDCs) have shown great potential as good quality thermoelectric materials at high temperature since past few years due to their suitable band gap tunabilty, low dimensionality and fantastic combination of electrical conductivity and lattice thermal conductivity. Here, a first principles calculations of electronic and thermoelectric performance of two dimensional monolayer $HfS_2$, $HfSe_2$ and their Janus monolayer HfSSe has been performed with the help of density functional theory and Boltzmann transport equation. Thermodynamical stability of all three structures has been confirmed from phonon dispersion curves. The thermoelectric parameters such as Seebeck coefficient, power factor and electrical conductivity have been calculated at 300K, 400K and 500K. The lattice thermal conductivity at room temperature has been found very low in monolayer $HfS_2$, $HfSe_2$ and HfSSe Janus monolayer as compared to very popular TMDCs such as $MoS_2$ and $WS_2$. An ultralow value of lattice thermal conductivity of the value of 0.36 W/mK at room temperature in Janus monolayer HfSSe has been found which is lower than that of monolayer $HfS_2$ and $HfSe_2$ because of the very low group velocity and short phonon lifetime in HfSSe. This ultralow lattice thermal conductivity in Janus monolayer HfSSe results a very high thermoelectric figure of merit close to the value of 1 at room temperature. By constructing the Janus monolayer of $HfS_2$ and $HfSe_2$ the thermoelectric performances significantly enhanced. Our theoretical investigation predicts that Janus monolayer HfSSe can be a revolutionary candidate for the fabrication of next generation wearable thermoelectric power generator to convert human body heat into electricity.

# Introduction:

Two-dimensional (2D) transition metal di-chalcogenides (TMDCs) have drawn very much attention since last few years in the field of electronics and optoelectronics due to their band gap tunability, atomically thick layer, unique density of states and mechanical strength[1]. Several experimental and theoretical studies reveal that they can be very promising materials with a suitable band gap beyond 2D Graphene for electronic[2], optoelectronic[3], thermoelectric[4], gas sensing[5], water splitting[6] and piezoelectric applications[7]. These TMDCs with a general formula of $MX_2$ (where M=Mo, W, Hf, Zr, Pt etc. and X=S, Se and Te), are layered materials with very weak Van-der waal interlayer interaction in which one single layer consists of one transition metal layer sandwiched between two chalcogen atoms layer. Recently Ang-Yu Lu et.al.[8] synthesized a new type of TMDCs by fully replacing the upper layer S atoms by Se atoms in monolayer MoS2 to form Janus monolayer MoSSe with the broken in-plane inversion symmetry and out of plane mirror symmetry. Inspiring from the experimental synthesis[8,9] of Janus monolayer MoSSe various studies on the electronic, optical, mechanical, thermoelectric, photocatylatic, gas sensing and mechanical properties of Janus TMDCs have been performed[10–17]. The absence of in-plane inversion symmetry and out of plane mirror symmetry results valley polarization and Rashba spin splitting in Janus monolayer TMDCs which make them suitable candidates in electronics and valleytronics[18].

Various experimental and theoretical studies reveal that 2D TMDCs are the promising candidates for thermoelectric application to convert wastage heat into electricity due to the unique combination of electrical conductivity and lattice thermal conductivity[19–21]. High thermoelectric power factor of the value of 8.5 mW m$^{-1}$K$^{-2}$ has been experimentally observed in few layer MoS$_2$ nanosheet motivating investigation on thermoelectric properties of other 2D TMDCs[22]. Theoretical predictions suggest the thermoelectric figure of merit (ZT) of these TMDCs are low as compared to the commercial available thermoelectric materials[23] Bi$_2$Te$_3$ and Bi$_2$Se$_3$ and various efforts such as chemical doping[24], strain engineering[2,25,26], making heterostructures[27–29] have been performed to increase the ZT product for efficient conversion of wastage heat into electricity. Monolayer ZrX$_2$ and HfX$_2$ where X= S, Se and Te, and their heterostructures[30–32] has greater potential in thermoelectric application with thermoelectric power factor close to unity because of their very low lattice thermal conductivity in comparison to that of well-known MoS$_2$ and WS$_2$ monolayer[33–36]. Experimental synthesis of ZrSe$_2$ and HfSe$_2$ and fabrication of devices with on/off > 10$^6$ reveals that they can be promising materials for low power electronic devices[37]. The electronic, optical and thermoelectric properties can be tuned by introducing structural defects and dopants and enhanced thermoelectric and optoelectronic properties can be observed in monolayer HfS$_2$[38]. The thermoelectric properties can be enhanced by the application of strain in monolayer ZrS$_2$[39] and ZrSe$_2$[40] because of the reduction in lattice thermal conductivity due to strain whereas thermoelectric properties of monolayer WS$_2$ can be enhanced with the application of compressive strain because of valley degeneracy[41]. Recent investigations on the thermoelectric properties of Janus monolayer MoSSe and ZrSSe predict that the thermoelectric performances can be effectively enhanced by forming Janus monolayer[42,43].

Though various studies have been performed on the thermoelectric performance of 2D TMDCs but there is still a lack of studies on the near room temperature thermoelectric materials for efficient conversation of heat into electricity near room temperature. Here, for the first time to best of our knowledge we have performed a detailed systematic investigation on the electronic and thermoelectric performance of Janaus monolayer HfSSe with the help of density functional theory and Boltzmann transport equation. In HfSSe Janus monolayer a ZT product close to unity has been found at 300K from our investigation which is greater than that of monolayer HfS$_2$ and HfSe$_2$. For the first time the lattice thermal conductivity of Janus HfSSe has been calculated in this paper with an ultralow value of 0.36 W/mK at 300K which results the high ZT factor in HfSSe at room temperature. Our results reveal the enhancement of thermoelectric performance due to reduction in lattice thermal conductivity because of lower group velocity and shorter phonon lifetime by constructing Janus monolayer HfSSe. Therefore, our theoretical investigation suggests that Janus monolayer HfSSe can be very promising thermoelectric materials near room temperature for the fabrication of next generation wearable thermoelectric power generator which can convert the heat from human body into electricity.

## Computational Details:

First principles calculations have been performed using density functional theory (DFT) with projector augmented wave (PAW)[44] potentials and the Perdew-Burke-Ernzerhof (PBE)[45] generalized gradient approximation (GGA)[46] as exchange correlation functional in Quantum Espresso (QE) package[47]. A sufficient vacuum of 17Å along C axis was created to avoid the interaction between layers in periodic boundary condition for all the monolayers. A 12×12×1 dense mesh grid was used to optimize the geometry and energy cutoff for the electronic wavefunctions was set to 50 Ry throughout all the calculations. For the density of state (DOS) a high dense mesh of k points 48×48×1 was used. For the calculation of thermoelectric parameters, we have constructed a 4×4×1 supercell where k points were sampled with 12×12×1 dense mesh grids. For the thermoelectric and transport properties, a semi classical Boltzmann transport theory was used with constant scattering time approximation (CSTA) as implemented in BoltzTrap[48] code. In CSTA we assume that scattering time is almost independent on energy and both the group velocity of carriers and DOS contribute to the transport function. The group velocity ($v_g$) of carriers in a specific band can be described as

$$v_\alpha(i, \mathbf{k}) = \frac{1}{\hbar} \frac{\partial \epsilon(i, \mathbf{k})}{\partial \mathbf{k}_\alpha}, \quad (1)$$

where $\mathbf{k}_\alpha$ is the α$^{th}$ component of wavevector $\mathbf{k}$ and $\epsilon(i, \mathbf{k})$ is the $i^{th}$ energy band and the conductivity tensor can be obtained in terms of group velocity as

$$\sigma_{\alpha\beta}(i, \mathbf{k}) = e^2 \tau(i, \mathbf{k}) v_\alpha(i, \mathbf{k}) v_\beta(i, \mathbf{k}), \quad (2)$$

The Seebeck coefficient, electrical conductivity and thermal conductivity due to electron can be calculated by using the values of group velocity $v_\alpha(i, \mathbf{k})$ as implemented in BoltzTrap[48] code by following equations

$$S_{\alpha\beta}(T, \mu) = \frac{1}{eT} \frac{\int v_\alpha(i, \mathbf{k}) v_\beta(i, \mathbf{k}) (\epsilon - \mu) \left[ -\frac{\partial f_\mu(T, \epsilon)}{\partial \epsilon} \right] d\epsilon}{\int v_\alpha(i, \mathbf{k}) v_\beta(i, \mathbf{k}) \left[ -\frac{\partial f_\mu(T, \epsilon)}{\partial \epsilon} \right] d\epsilon}, \quad (3)$$

$$\frac{\sigma_{\alpha\beta}(T, \mu)}{\tau(i, \mathbf{k})} = \frac{1}{V} \int e^2 v_\alpha(i, \mathbf{k}) v_\beta(i, \mathbf{k}) \left[ -\frac{\partial f_\mu(T, \epsilon)}{\partial \epsilon} \right] d\epsilon, \quad (4)$$

$$\frac{k_{\alpha\beta}^{el}(T,\mu)}{\tau(i,\boldsymbol{k})} = \frac{1}{TV}\int v_\alpha(i,\boldsymbol{k})v_\beta(i,\boldsymbol{k})(\epsilon-\mu)^2\left[-\frac{\partial f_\mu(T,\epsilon)}{\partial \epsilon}\right]d\epsilon, \quad (5)$$

where $e, T, \tau, \mu, V$ are electronic charge, temperature, relaxation time, chemical potential, volume of an unit cell respectively and $f_\mu(T,\epsilon) = \frac{1}{e^{(\epsilon-\mu)/K_BT}+1}$ is the Fermi-Dirac distribution function.

The lattice thermal conductivity ($k_{ph}$) has been calculated with the linearized phonon Boltzmann equation as implemented in Phono3py[49] code interfaced with Quantum Espresso (QE) package[47].

The thermoelectric figure of merit (ZT) has been calculated using the formula

$$ZT = \frac{S^2\sigma T}{k_{el}+k_{ph}}, \quad (6)$$

Where S, σ, T are Seebeck coefficient, electrical conductivity and temperature respectively and $k_{el}$ and $k_{ph}$ arethermal conductivity due to electron, and lattice thermal conductivity due to phonon respectively.

## Results and Discussions:

### Structural Parameters and structural stability:

The bulk HfX$_2$ (X=S, Se) has a 1T-CdI$_2$ type structure belonging to P3¯m1 space group. Due to very weak interlayer Vander Waals interaction monolayer can be obtained easily by exfoliation. In a monolayer the Hf atoms are sandwiched between two layers of S or Se atoms. The top view of the HfS$_2$ and HfSe$_2$ monolayer unit cell are shown in Fig. 1a and 1b. The optimized lattice constant of HfS$_2$ has been calculated as a=b=3.65Å and that of HfSe$_2$, a=b=3.78 Å for a unit cell of monolayer agrees with previous calculations[50] and experimental[51] value. The Janus monolayer HfSSe has been constructed by fully replacing one of the two S layers with Se atoms in HfS$_2$ monolayer. The Janus monolayer HfSSe belongs to P3m1 space group with slightly lower symmetry because of losing the reflection symmetry with respect to the central metal (Hf) atoms. The top and side view of HfSSe Janus monolayer unit cell has been shown in Fig. 1c. and 1d. The optimized lattice constant of Janus monolayer HfSSe has been calculated as a=b=3.71Å which is in between that of HfS$_2$ and HfSe$_2$ monolayers. The calculated lattice constant, bond lengths and bond angles for all the structures are listed in Table 1. It is observed that the Hf-S (or Se) bond length in $HfS_2$ (or $HfSe_2$) monolayers is almost same in HfSSe Janus monolayer whereas the S-Se bond length in HfSSe Janus monolayer is in between that of S-S (Se-Se) bond length in $HfS_2$ ($HfSe_2$) monolayers.

To check the dynamical stability of our structures we have calculated phonon dispersion curve of $HfS_2$, $HfSe_2$ and HfSSe Janus monolayer as shown in Fig. 2. It is clearly observed that there is no imaginary frequency on the negative side in the phonon dispersion curves of all the three structures indicating dynamical stability of our system. There are 3 acoustic and 6 optical modes observed in each dispersion curve because of the unit cell contains 3 atoms in each structure. Three lower frequency branches correspond to acoustic vibrational modes represented by in plane longitudinal acoustic (LA) and transverse acoustic (TA) mode and out of plane modes (ZA). The six upper frequency branches are optical modes. The in-plane LA and TA modes show linear relationship with k vector near the Γ point (k=0) but out of plane ZA branch has a quadratic relationship with k.

**Electronic band structures and density of states:**

The electronic band structures of monolayer $HfS_2$, $HfSe_2$ and Janus monolayer HfSSe have been calculated along the Brillouin zone path K-Γ-M-K as shown in Fig.3. The valance band maxima (VBM) lie at the high symmetry point Γ whereas the conduction band minima (CBM) lie at the high symmetry point M for all three monolayers ($HfS_2$, $HfSe_2$ and HfSSe) resulting an indirect band gap. The calculated values of indirect band gap are 1.30eV and 0.65eV for $HfS_2$ and $HfSe_2$ monolayer respectively which matches well with the previous calculated values[50]. For the Janus monolayer HfSSe the indirect band gap value has been calculated 0.834 eV which is in between that of monolayer $HfS_2$ and $HfSe_2$ band gap values.

The VBM is mainly contributed by $p_x$ and $p_y$ orbitals of S atoms and the CBM is mainly due to the contribution of $d_{z2}$ and $d_{zy}$ of Hf atoms and slightly contribution of $p_z$ orbitals of S atoms in $HfS_2$ monolayers as shown in Fig.4.a. Similarly, in $HfSe_2$ monolayer $p_x$ and $p_y$ orbitals of Se atoms contributes mainly to the VBM and $d_{z2}$ and $d_{zy}$ of Hf atoms with combination of $p_z$ orbitals of S atoms contribute mainly to the CBM as shown in Fig.4.b. Also, the gap between VBM and CBM is much larger for $HfS_2$ than that of $HfSe_2$ monolayers as $HfS_2$ has much higher band gap than $HfSe_2$ monolayer. There is negligible contribution of Hf atoms in valance band.

In case of HfSSe Janus monolayer the VBM is mainly contributed by Se $p_y$, Se $p_x$, S $p_y$ and S $p_x$ orbitals among which contribution of Se $p_y$ is more and it is clearly seen that the contribution of Se atoms is more than that of S atoms to the DOS as shown in Fig.4.c. There is no contribution of $p_z$ orbitals of S or Se atoms near the band edges. So in Janus monolayer HfSSe Se atoms dominates more to the energy states than that of S atoms. The CBM is mainly contributed by $d_{z^2}$, $d_{zx}$ and $d_{xy}$ orbitals of Hf atoms and $p_z$ orbitals of S and Se atoms in Janus monolayer HfSSe. Unlike $HfS_2$ and $HfSe_2$, in HfSSe $p_x$ and $p_y$ orbitals of S and Se atoms are not degenerate whereas that of $HfS_2$ and $HfSe_2$ were degenerate near the band edges. Also, in HfSSe Janus monolayer the gap between VBM and CBM is larger than that of HfSe2 but lower than that of HfS2 monolayer. So, it is seen that in Janus monolayer HfSSe, Se atoms dominate more than S atoms in valance band and in conduction band Hf atoms dominate more.

## Thermoelectric Properties:

The variation of Seebeck coefficient (S) with chemical potential (µ) at 300K, 400K and 500K in $HfS_2$ monolayer is shown in Fig. 5.a. The highest Seebeck coefficient is found to be 2148 µV/K for n-type carriers (µ>0) and that of 2005 µV/K for p-type carriers (µ<0) at 300K for HfS2 monolayer. As temperature increases the value of S decreases. The variation of electrical conductivity($\sigma/\tau$) and relaxation time scaled power factor($S^2\sigma/\tau$) with chemical potential (µ) is shown in Fig. 5.b. and 5.c. respectively. The highest power factor with a value of 19.73 $\times10^{10}$ W/m$^2$Ks for n-type carriers and 12.81$\times10^{10}$ W/m$^2$Ks for p-type carriers at 500K indicates that n-type doping is more efficient than p-type doping in monolayer $HfS_2$ for thermoelectric application. The variation of S, $\sigma/\tau$ and $S^2\sigma/\tau$ with chemical potential (µ) in monolayer $HfSe_2$ and Janus monolayer HfSSe at different temperatures are shown in Fig.6. and Fig.7 respectively. The highest values of S, $\sigma/\tau$ and $S^2\sigma/\tau$ for both n-type and p-type carriers in HfS2, HfSe2 and HfSSe Janus monolayer at 300K, 400K and 500K are listed in Table 2. It is observed that monolayer $HfS_2$ has highest Seebeck coefficient and monolayer $HfSe_2$ has lowest Seebeck coefcient where as in Janus monolayer HfSSe, Seebeck coefficient has a value in between. This is analogous to the fact that Seebeck coefficient depends on the band gap value and $HfS_2$ has highest bandgap and $HfSe_2$ has lowest bandgap among these three where Janus monolayer HfSSe has a band gap value in between that of $HfS_2$ and $HfSe_2$ monolayer as discussed earlier. Also, for all three structures it is clearly seen that thermoelectric properties for n-type carriers are better than that of p-type carriers concluding higher effectiveness of n-type doping than p-type doping for thermoelectric application.

**Lattice thermal conductivity:**

The variation of lattice thermal conductivity due to phonon with temperature in monolayer $HfS_2$, $HfSe_2$ and HfSSe Janus monolayer is shown in Fig.8. a. Monolayer $HfS_2$ and $HfSe_2$ has very low lattice thermal conductivity as compare to well-known 2D TMDCs $MoS_2$ and $WS_2$. Monolayer $HfS_2$ and $HfSe_2$ has a lattice thermal conductivity of the value of 2.836 W/mK and 1.517 W/mK respectively at 300K which matches well with the previously reported values[33,52]. A significant reduction in lattice thermal conductivity in Janus monolayer HSSe has been observed. The lattice thermal conductivity in Janus monolayer HfSSe has been found to be 0.36 W/mK at 300K, which is ultralow as compared to $HfS_2$ and $HfSe_2$ monolayer. For all three structures lattice thermal conductivity decreases with increasing temperature. The reason of such low values of lattice thermal conductivity can found from phonon dispersion curve of these materials as shown in Fig.2.a. From the phonon dispersion curves it is clearly observed that there is no coupling between acoustic mode LA and lower optical mode in $HfS_2$ monolayer but acoustic LA mode is coupled to the lower optical mode in $HfSe_2$ and HfSSe. Due to that phonon-phonon coupling there is reduction in phonon group velocity and scattering increases, as a result lattice thermal conductivity due to phonon reduced significantly decreases in $HfSe_2$ and HfSSe monolayer as compared to that of $HfS_2$ monolayer. The percentage contribution of different modes to the lattice thermal conductivity in $HfS_2$, $HfSe_2$ and HfSSe has been shown in Fig.8. b., Fig 8.c. and Fig 8.d respectively. For all three structures lattice thermal conductivity is mainly contributed by acoustic modes and there is very small contribution of optical modes. In monolayer $HfS_2$ the LA mode contributes almost 40% of the total value whereas ZA and TA modes contribute 31% and 23% respectively. In monolayer $HfSe_2$, almost half part of the lattice thermal conductivity is contributed by LA mode and the contribution of TA modes remains almost same as that of $HfS_2$ but contribution of the out of plane ZA mode decreases to 10% in $HfSe_2$. In HfSSe Janus monolayer the two higher frequency acoustic modes LA and TA contribute almost same (greater than 30%) whereas the ZA modes contributes less than 10% of the total lattice thermal conductivity.

## Analysis of Ultra-low lattice thermal conductivity:

The ultralow value of lattice thermal conductivity of Janus monolayer HfSSe can be explained from the group velocity and phonon relaxation time. The variation of group velocity with frequency in monolayer $HfS_2$, $HfSe_2$ and Janus monolayer HfSSe is shown in Fig.9. From this figure it is clearly seen that the phonons in HfSSe and $HfSe_2$ has lower group velocity than that of $HfS_2$ as a result heat will be transferred slowly in HfSSe and $HfSe_2$ subjected to the lower value of lattice thermal conductivity. Also, phonon lifetime is one of the important factors in transport of heat energy. The variation of phonon lifetime with frequency in monolayer $HfS_2$, $HfSe_2$ and Janus monolayer HfSSe is shown in Fig.10. The phonons in monolayer $HfS_2$ have highest lifetime around 4 ps at lower frequency range where in $HfSe_2$ phonons lifetime is around 2.5 ps which is shorter than that of $HfS_2$. In Janus monolayer HfSSe, a very significant reduction in phonon lifetime has been observed as shown in Fig.10. c. In HfSSe the phonons have shortest lifetime of the value of 0.6 ps which is much less than that of $HfS_2$ and $HfSe_2$. Due to the very short lifetime of phonons in Janus monolayer HfSSe, there will be a lot of scattering which results suppression of phonons in the system and that is the reason of the ultralow value of lattice thermal conductivity in Janus monolayer HfSSe. Such type of very low values of lattice thermal conductivity predicts that $HfS_2$, $HfSe_2$ and Janus monolayer HfSSe can be used as a very promising room temperature thermoelectric material.

## The thermoelectric figure of merit:

The quantity thermoelectric figure of merit (ZT) defines the quality and efficiency of a thermoelectric material. For obtaining optimal ZT product the electrical conductivity should be high and the lattice thermal conductivity should be very low. The ZT product is defined as

$$ZT = \frac{S^2 \sigma T}{k_{el} + k_{ph}},$$

where S, σ, T are Seebeck coefficient, electrical conductivity and temperature respectively and $k_{el}$ and $k_{ph}$ are thermal conductivity due to electron, and lattice thermal conductivity due to phonon respectively. The variation of the calculated ZT product as a function of chemical potential (μ) of monolayer $HfS_2$, $HfSe_2$ and Janus HfSSe at 300K has been shown in Fig.11. a. The Janus monolayer HfSSe has highest ZT product of the value of 0.86 for n-type carriers and 0.70 for p-type carriers. This high value of ZT product at room temperature in Janus monolayer HfSSe is due to its ultralow lattice thermal conductivity. The calculated ZT product in HfS$_2$ and HfSe$_2$ has been found to be 0.65 and 0.70 for n-type carriers and 0.49 and 0.54 for p-type carriers respectively. The highest ZT product of all three structures has been calculated at three different temperatures 330K, 400K and 500K are plotted in Fig.11.b. for n-type carriers and Fig.11.c. for p-type carriers. It is observed that ZT product for n-type carriers is higher than that of p-type carriers for all the three structures implies higher effectiveness of n-type doping than that of p-type doping. The room temperature ZT product in HfX$_2$(X=S,Se) has been found greater than 0.65 for n-type carriers which is much higher than that of very popular 2D TMDCs MoS$_2$ and WS$_2$ and that implies that monolayer HfX$_2$ can be promising materials as room temperature thermoelectric materials. The highest ZT product in Janus monolayer HfSSe has been found to be 0.93 for n-type and 0.86 for p-type carriers at 500K. It is clearly seen that the thermoelectric figure of merit (ZT) enhances significantly because of the construction of Janus monolayer HfSSe. In Janus monolayer the power factor ($S^2\sigma/\tau$) remains almost same as that of monolayer HfS$_2$ and HfSe$_2$ but the lattice thermal conductivity decreases significantly and due to the ultralow lattice thermal conductivity the thermoelectric performance increases efficiently in Janus monolayer HfSSe. So, construction of Janus monolayer HfSSe by fully replacing one layer with S or Se atoms, can be an effective way to enhance the thermoelectric performance of this type of materials and achieving high ZT product close to the value of 1 at a temperature of 500K.

## Conclusions:

In conclusion, we have calculated the electronic and thermoelectric properties of monolayer HfS$_2$, HfSe$_2$ and Janus monolayer HfSSe by fully replacing one layer of S in HfS$_2$ by Se atoms, by using DFT and Boltzmann transport theory. The dynamical stability of Janis monolayer has been checked by phonon dispersion curve and no imaginary frequency has been found in negative region implies our Janus monolayer HfSSe is stable. Though there is very small change in thermoelectric power factor ($S^2\sigma/\tau$) in Janus HfSSe than that of HfS$_2$ and HfSe$_2$ but a significant reduction in lattice thermal conductivity has been observed because of the lower group velocity and shorter phonon lifetime in Janus monolayer HfSSe. Due to the ultralow lattice thermal conductivity of the value of 0.36 W/mK at 300K in Janus monolayer HfSSe the ZT product has been found to be 0.86 for n-type carriers and 0.70 for p-type carriers at room temperature. However, the highest ZT product of the value of 0.93 for n-type and 0.86 for p-type carriers in Janus monolayer HfSSe at 500K which are very close to unity. It is clear that thermoelectric performance enhances significantly in Janus monolayer HfSSe than that of monolayer $HfS_2$ and $HfSe_2$. So, construction of Janus monolayer can be an effective way to enhance the thermoelectric performance of two dimensional TMDCs. Our theoretical study on thermoelectric properties of Janus monolayer HfSSe implies that it could be a fantastic 2D thermoelectric material for the fabrication of next generation high-performance wearable thermoelectric power generator to convert human body heat into electricity.

## Acknowledgement:

The authors are thankful to Ministry of Human Resource and Development (MHRD) India for supporting this work. We are also thankful to Indian Institute of Technology Jodhpur (IITJ) for all the supports to carry out the experiment.

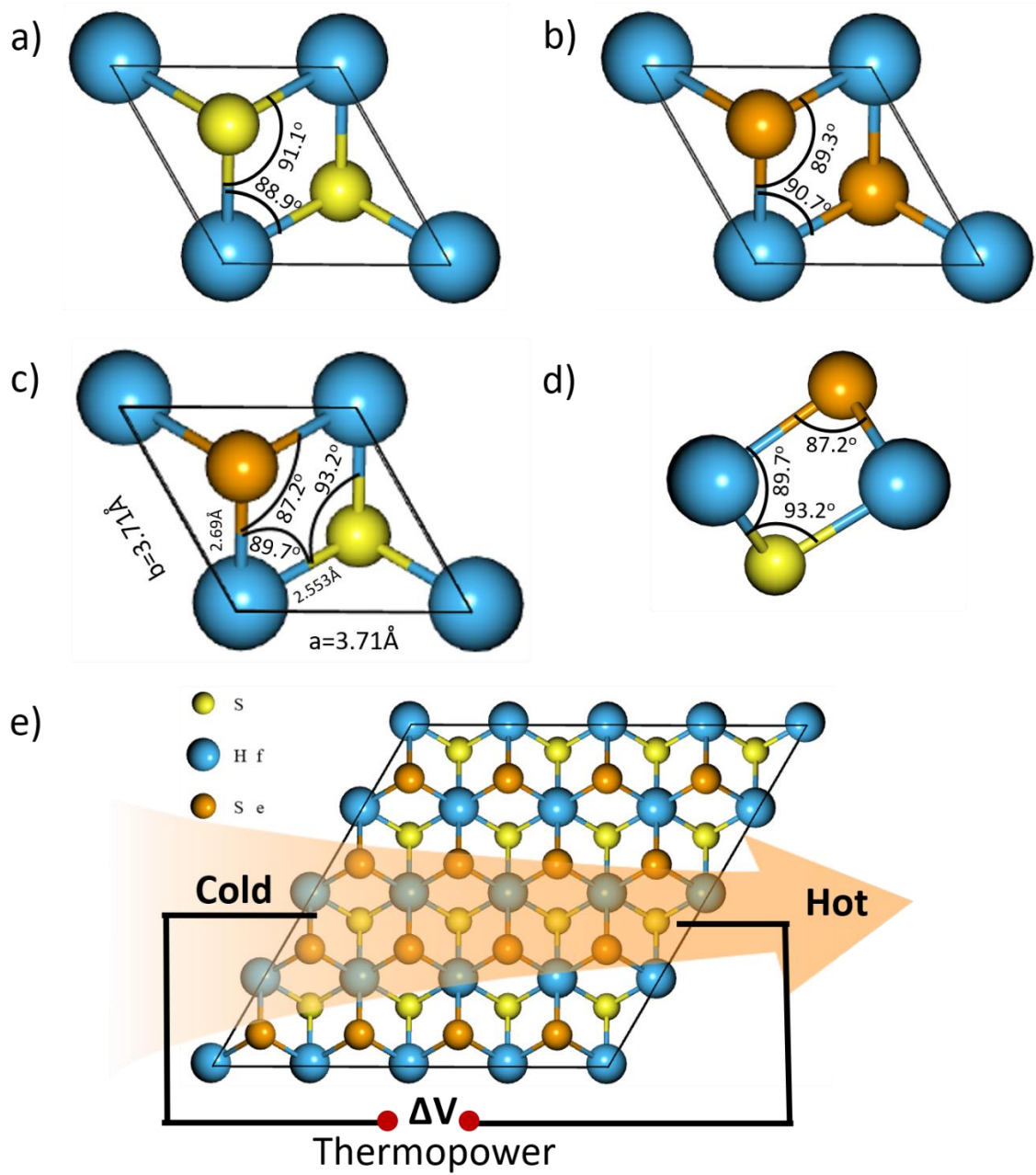

**Fig.1.** Crystal structure of the a) HfS$_2$ unit cell b) HfSe$_2$ unit cell c) HfSSe Janus monolayer unit cell top view and d) side view e) top view of a 4×4×1 supercell for calculation of thermoelectric properties.

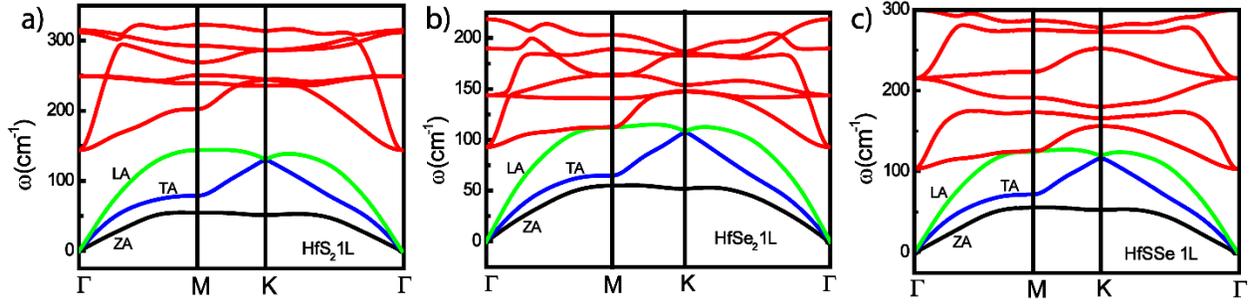

**Fig.2.** Phonon dispersion curve of the monolayer a) $HfS_2$ b) $HfSe_2$ and c) HfSSe Janus monolayer.

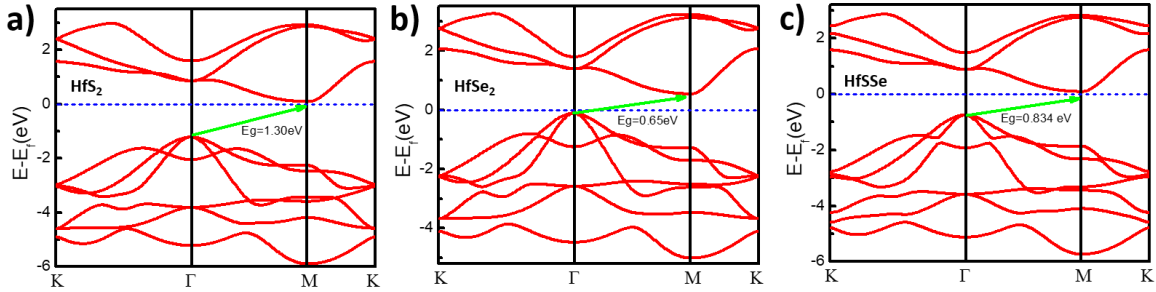

**Fig.3.** Electronic band structure of monolayer a) $HfS_2$ b) $HfSe_2$ and c) HfSSe Janus monolayer along K-Γ-M-K path. The green arrows show indirect band gap and the Fermi level is represented by blue dotted lines.

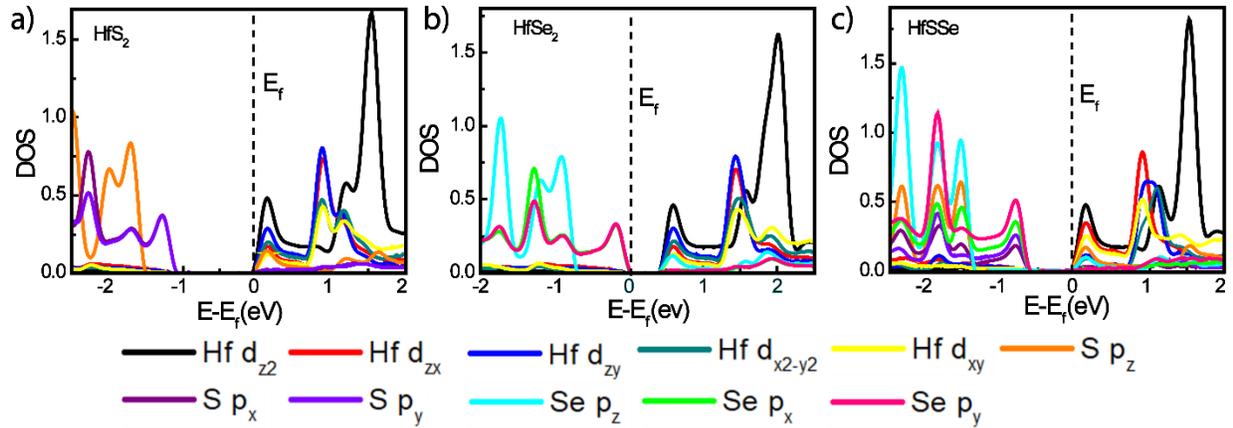

**Fig.4.** Projected density of states (PDOS) of monolayer a) $HfS_2$ b) $HfSe_2$ and c) HfSSe Janus monolayer. Fermi level is shown by black dotted lines. Different colours represent different orbitals of Hf, S and Se atoms.

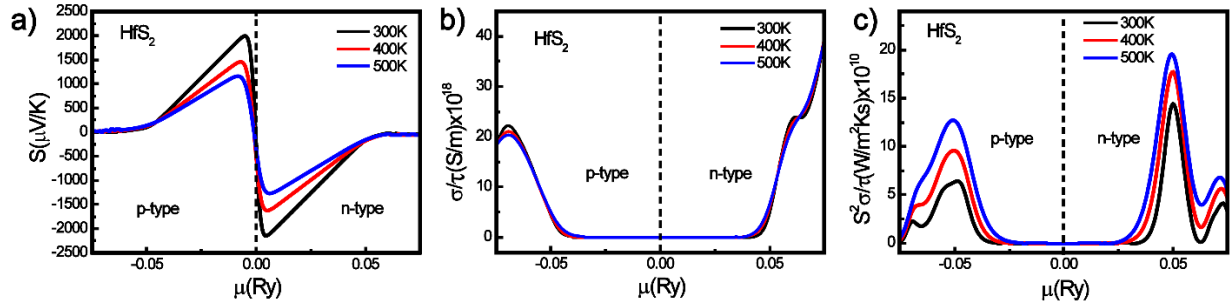

**Fig.5.** The variation of a) Seebeck coefficient (S) b) electrical conductivity ($\sigma/\tau$) and c) relaxation time scaled thermoelectric power factor ($S^2\sigma/\tau$) as a function of chemical potential ($\mu$) at 300K, 400K and 500K in HfS$_2$ monolayer

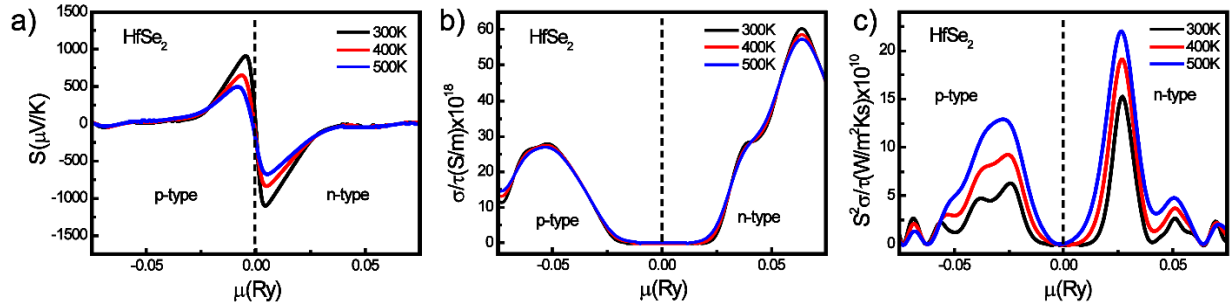

**Fig.6.** The variation of a) Seebeck coefficient (S) b) electrical conductivity ($\sigma/\tau$) and c) relaxation time scaled thermoelectric power factor ($S^2\sigma/\tau$) as a function of chemical potential ($\mu$) at 300K, 400K and 500K in HfSe$_2$ monolayer

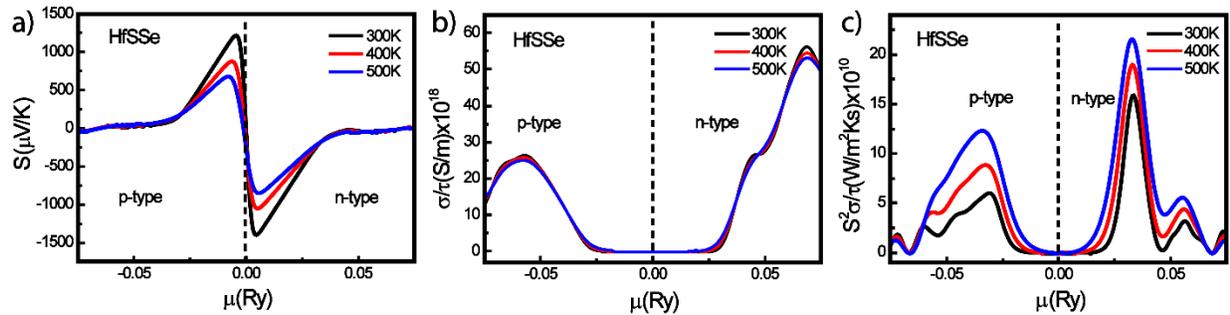

**Fig.7.** The variation of a) Seebeck coefficient (S) b) electrical conductivity ($\sigma/\tau$) and c) relaxation time scaled thermoelectric power factor ($S^2\sigma/\tau$) as a function of chemical potential ($\mu$) at 300K, 400K and 500K in Janus monolayer HfSSe

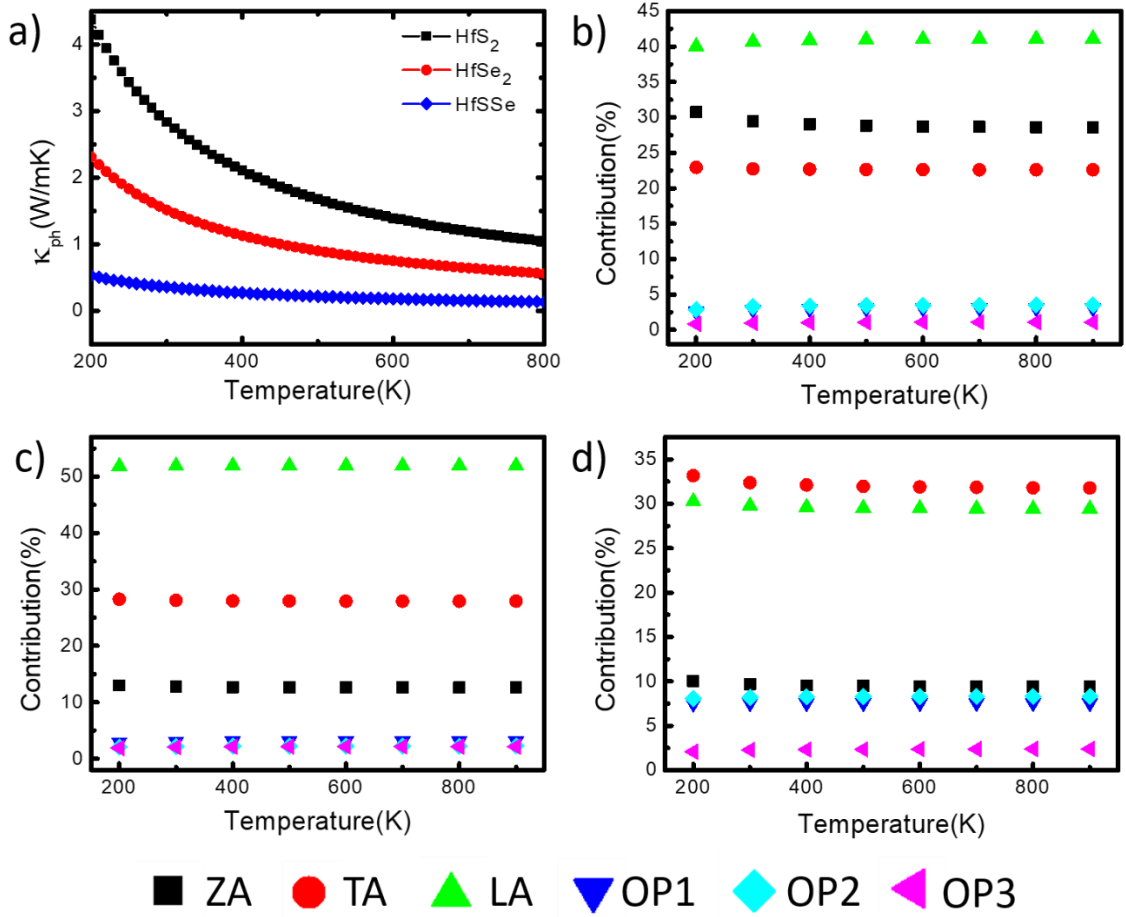

**Fig. 8.** a) The variation of lattice thermal conductivity due to phonon with temperature, percentage contribution of different acoustic and optical modes in lattice thermal conductivity in monolayer b) HfS2 c) HfSe2 and d) HfSSe with temperature.

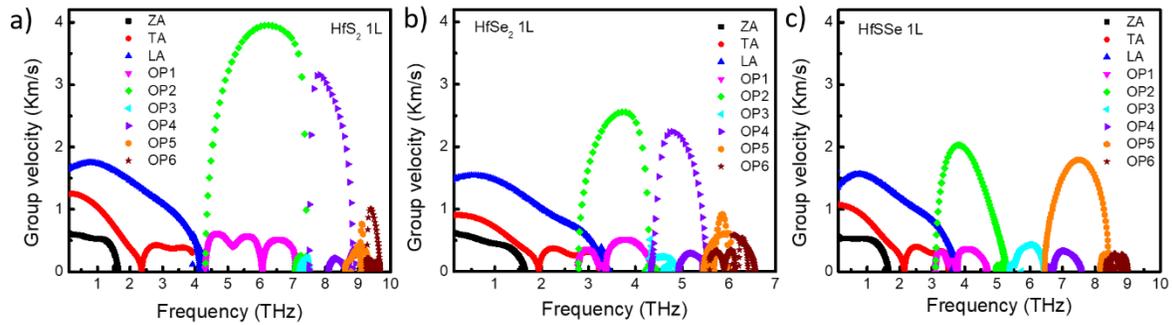

**Fig. 9.** Phonon group velocity of all nine modes as a function of frequency in monolayer a) HfS2, b) HfSe2 and c) HfSSe Janus monolayer.

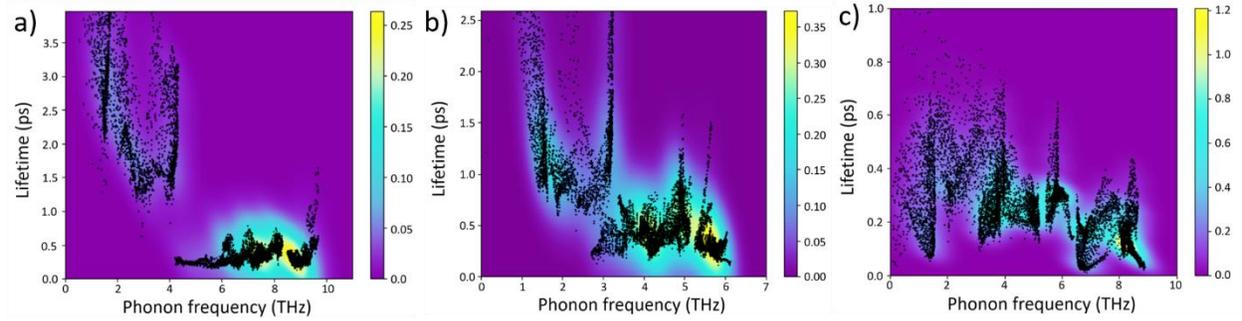

**Fig. 10.** The phonon lifetime as a function of frequency in monolayer a) HfS2, b) HfSe2 and c) HfSSe Janus monolayer. It is clearly seen that HfSSe Janus monolayer has smallest phonon lifetime among all the three structures owing to smallest lattice thermal conductivity.

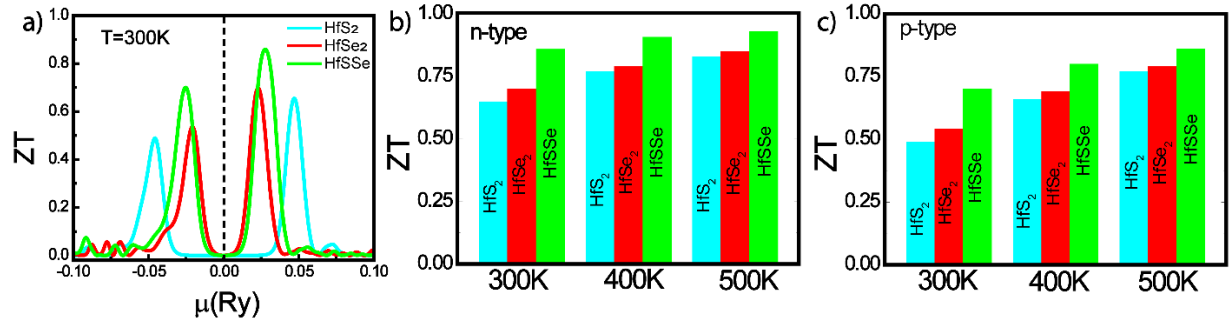

**Fig. 11.** The variation in a) thermoelectric figure of merit (ZT) as a function of chemical potential(μ) at 300K in monolayer HfS2, HfSe2 and HfSSe Janus monolayer. The bar plot of the highest ZT product at 300K, 400K and 500K in HfS2, HfSe2 and HfSSe for b) n-type and c) p-type carriers.

|  | a=b (Å) | $d_{Hf-S}$ (Å) | $d_{Hf-Se}$ (Å) | $d_{S/Se-S/Se}$ (Å) | <Hf-S-Hf | <Hf-Se-Hf | <S/Se-Hf-S/Se |
|---|---|---|---|---|---|---|---|
| $HfS_2$ | 3.65 | 2.55 | - | 3.58 | 91.1° | - | 88.9° |
| $HfSe_2$ | 3.78 | - | 2.69 | 3.83 | - | 89.3° | 90.7° |
| HfSSe | 3.71 | 2.55 | 2.69 | 3.70 | 93.2° | 87.2° | 89.7° |

**Table 1.** Calculated lattice parameter(a=b), different bond lengths and bond angles of monolayer HfS$_2$, HfSe$_2$ and Janus monolayer HfSSe.